\documentstyle[11pt,newpasp,twoside,epsf]{article}
\reversemarginpar

\begin{document}
\title{High Sensitivity, High Spectral Resolution, Mid-infrared Spectroscopy}
\author{Matthew J. Richter, John H. Lacy, Daniel T. Jaffe, Thomas K. Greathouse, Qingfeng Zhu}
\affil{University of Texas Astronomy Department, Austin, TX 78712}

\begin{abstract}
We broadly discuss mid-infrared spectroscopy and detail our new
high spectral resolution instrument, the Texas Echelon-cross-Echelle
Spectrograph (TEXES).  
\end{abstract}

\section{Introduction}
In Bob Tull's design and construction of high resolution, optical
spectrographs at McDonald Observatory, he has had the advantage of
being able to crawl around inside his spectrograph to see 
how the instrument is behaving.  That advantage, along with careful
design and great experience, has allowed him to make spectrographs like
2dcoude: R$\equiv {\lambda\over{\Delta\lambda}}$ up
to 250,000 or R=60,000 and complete optical coverage in two settings
(Tull et al. 1995).
We have built a spectrograph that does not quite equal
Bob's work in terms of maximum spectral resolution or fractional wavelength
coverage, but operates at 20 times longer wavelength with 60 times 
fewer pixels. It can rightfully
be called the first true high spectral resolution grating 
spectrograph for the mid-infrared (MIR).

The MIR, here defined as 5~\micron$<\lambda<$25~\micron, is 
somewhat exotic.  Section 2
gives some background about 
observing near 10~\micron.  Section 3 describes
MIR spectral features, past results, and 
present and future instruments.  In Section 4
we discuss methods of obtaining R=100,000 at 10~\micron.
Finally, in Section 5, we describe our instrument, TEXES, in more
detail and present some preliminary results.

\section{Observing in the MIR}
Ground-based MIR astronomy is predominantly limited by the atmosphere
and background radiation.  The Earth's atmosphere limits observations
to spectral regions, windows, where terrestrial molecules do not absorb
light from space.  More importantly, molecules that absorb photons also 
emit photons,
contributing to the background photon rate.  In most cases, photon noise
from the background emission is the limiting factor in the sensitivity
of MIR instruments.

Figure 1 shows the MIR atmospheric transmission from Mauna Kea on a dry 
night.
\begin{figure}
\plotfiddle{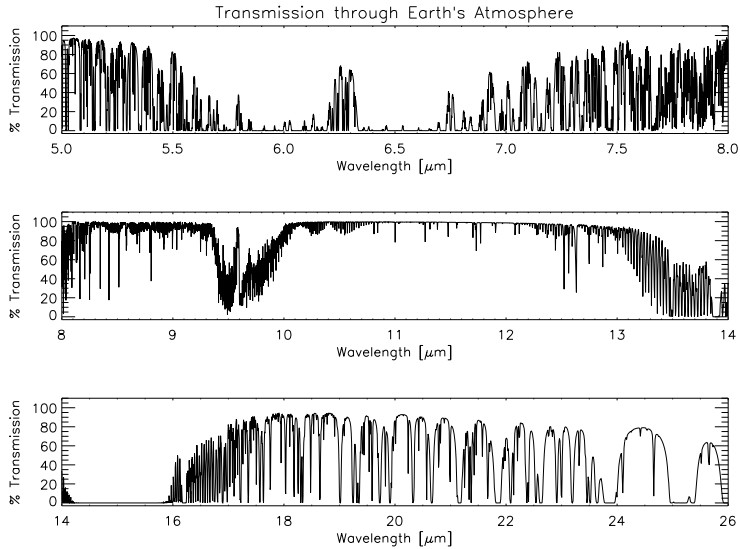}{2.75in}{0}{45}{45}{-150}{0}
\caption{Transmission through the Earth's atmosphere based on a model
calculation.  Some of the major molecular absorbers are identified in
the text.}
\end{figure}
The dominant absorbers 
are H$_2$O (5.5-7.5~\micron, 17-30~\micron), CO$_2$
(13.5-16.5~\micron), and O$_3$ (9.5-10~\micron).  Water is particularly
devastating because the lines form low in Earth's atmosphere and are
substantially pressure-broadened.  Observing
from high, dry sites substantially improves the transmission (Figure 2)
\begin{figure}
\plotfiddle{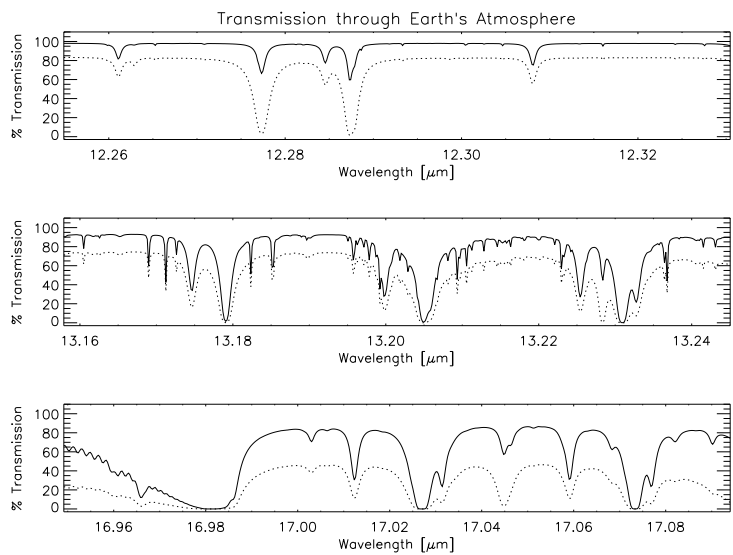}{2.75in}{0}{45}{45}{-150}{0}
\caption{Models of the terrestrial transmission for
elevations of 14,000~feet (solid line) and 7,000~feet (dotted line).
The far wings of water lines are responsible for the reduction in the
peak transmission at lower altitudes.}
\end{figure}

The background photon rate is given by
\begin{displaymath}
N_\gamma = 2.36 \times 10^{-11} {\epsilon B_\nu(T)\over h \nu } A\,\,\, {\rm 
[photons/s/Hz/arcsec^2]}
\end{displaymath}
where $\epsilon$ is the sky+telescope+instrument emissivity, $B_\nu(T)$ is
the Planck function, $A$ is the telescope area, $h\nu$ is the energy of
the photon, and the numerical factor converts the solid angle to $arcsec^2$.
A typical number for a 3~meter telescope with $\epsilon=0.1$ would be 
5$\times 10^9$~photons~s$^{-1}$~\micron$^{-1}$~arcsec$^{-2}$ 
or a background 
of approximately -2.7~mag~arcsec$^{-2}$.
Note that a cooled, space telescope will have orders of magnitude improvement
in sensitivity.

\section{MIR Spectroscopy}

MIR spectral features can be loosely divided into three categories:
solid features, ionic lines, and molecular rotation-vibration transitions.
Each of these classes probes a different environment and is best observed
with different spectral resolution.
Solid features include dust grains, ices, and polycyclic aromatic hydrocarbons
(PAHs).  The broad nature of these features suggest observations at 
low spectral resolution, R$\sim$100~to~1000.
Ionic and atomic spectral lines, for example from HII regions, 
external galaxies, or shock excited regions, require higher resolving power:
R$\sim$1000~to~10,000.  Molecular transitions are often associated with
cold interstellar gas or stellar photospheres.  Observing molecular lines 
requires high spectral resolution: R$\sim$10,000~to~100,000.

Most molecular transitions in the MIR are rotation-vibration
bands.  Therefore, many lines, each arising from a different
rotational energy level of the same molecule, are closely spaced
in frequency.  In addition, isotopic transitions and transitions
involving higher vibrational states may also overlap in frequency.
With sufficient spectral coverage one can simultaneously observe
transitions with a range of sensitivity to temperature and
density, providing a consistent data set with multiple constraints.

Table 1 lists commonly observed molecules in the MIR.
Molecules without an electric dipole, H$_2$,
CH$_4$, and C$_2$H$_2$, are unobservable with radio telescopes.

\begin{table}
\begin{center}
\caption{Important MIR molecules}
\begin{tabular}{ll}
\tableline
Molecule & Comments \\
\tableline
H$_2$ & pure rotational lines throughout MIR \\
H$_2$O & pure rotation and ro-vibrational lines throughout MIR \\
CH$_4$ & band at 7.7~\micron \\
C$_2$H$_2$ & band at 13.7~\micron \\
HCN & band at 14.0~\micron \\
C$_2$H$_6$ & bands at 12.2 \\
SiO & band at 8.1~\micron \\
NH$_3$ & bands at 10~\micron \\
\tableline
\tableline
\end{tabular}
\end{center}
\end{table}

Table 2 provides a non-exhaustive list of some past, present, and
future MIR spectrographs along with their resolving power.  
Clearly, very few instruments are designed to concentrate
on MIR molecular spectroscopy.  Unfortunately, infrared space telescopes,
although extremely sensitive,
are not equipped with high resolution spectrographs primarily because of
weight and size limitations.

\begin{table}
\caption{Some Past, Present and Future MIR Spectrographs}
\begin{center}
\begin{tabular}{lll}
\tableline
Instrument & Resolving Power & Telescope  \\
\tableline
FTS & 50,000 & KPNO (past) \\
Irshell & 10,000 & IRTF and ESO (past) \\
ISO SWS & 2,000 (10,000) & ISO satellite (past) \\
Keck LWS & 1,000 & Keck (present) \\
Celeste & 10,000 & KPNO and IRTF (present) \\
HIPWAC & 1,000,000 & IRTF (present) \\
TEXES & 100,000 & IRTF (present) \\
COMICS & 1-5,000 & Suburu (present) \\
Michelle & 1-30,000 & UKIRT and Gemini (future) \\
VISIR & 1-30,000 & VLT (future) \\
SIRTF IRS & 600 & SIRTF satellite (future) \\
\tableline
\tableline

\end{tabular}
\end{center}
\end{table}

MIR spectroscopy definitely advanced in importance with the 
success of the Infrared Space Observatory (ISO).  ISO's
Short Wavelength Spectrometer (SWS; de Graauw et al. 1996), provided
many beautiful data sets throughout the MIR with a resolving power $\sim$2000.
Figures 3 and 4 present examples of SWS data.  Figure 3 shows
solid features and some ionic lines from evolved stars (Molster et al. 1996).
\begin{figure}
\plotfiddle{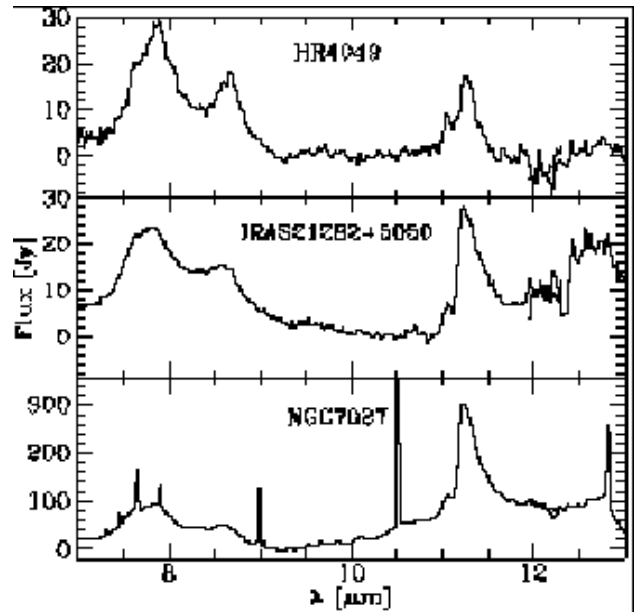}{2.75in}{0}{70}{70}{-210}{-160}
\caption{Taken from Molster et al. (1996) showing the MIR spectra of several
evolved stars.  The broad features are PAHs.  The sharp lines in the bottom
panel are ionic emission lines.}
\end{figure}
Figure 4 shows Jupiter in the MIR along with a model spectrum and
identification of molecular features (Encrenaz et al. 1996).  
\begin{figure}
\plotfiddle{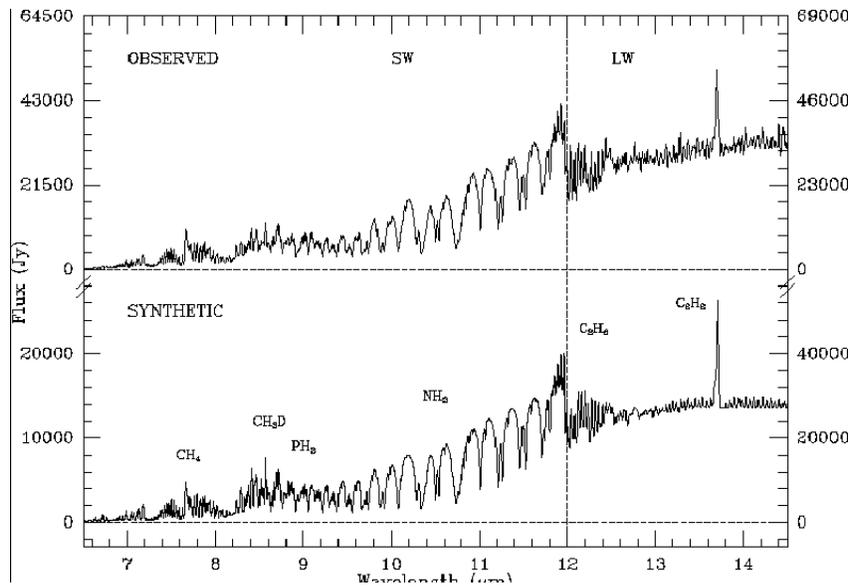}{2.75in}{270}{40}{40}{-150}{250}
\caption{SWS spectrum of Jupiter plus model identifying some molecular bands.
(Encrenaz et al. 1996).}
\end{figure}
Both figures show the tremendous
signal-to-noise and wavelength coverage achieved by the SWS.
However, it is important to note that with the SWS spectral resolution,
narrow features such as the C$_2$H$_2$ Q-branch at 729~cm$^{-1}$ are
totally blended together.

Figure 5 is a figure taken from Evans et al. (1991) showing the
observations of the C$_2$H$_2$ Q-branch in absorption toward IRc2 in Orion.
\begin{figure}
\plotfiddle{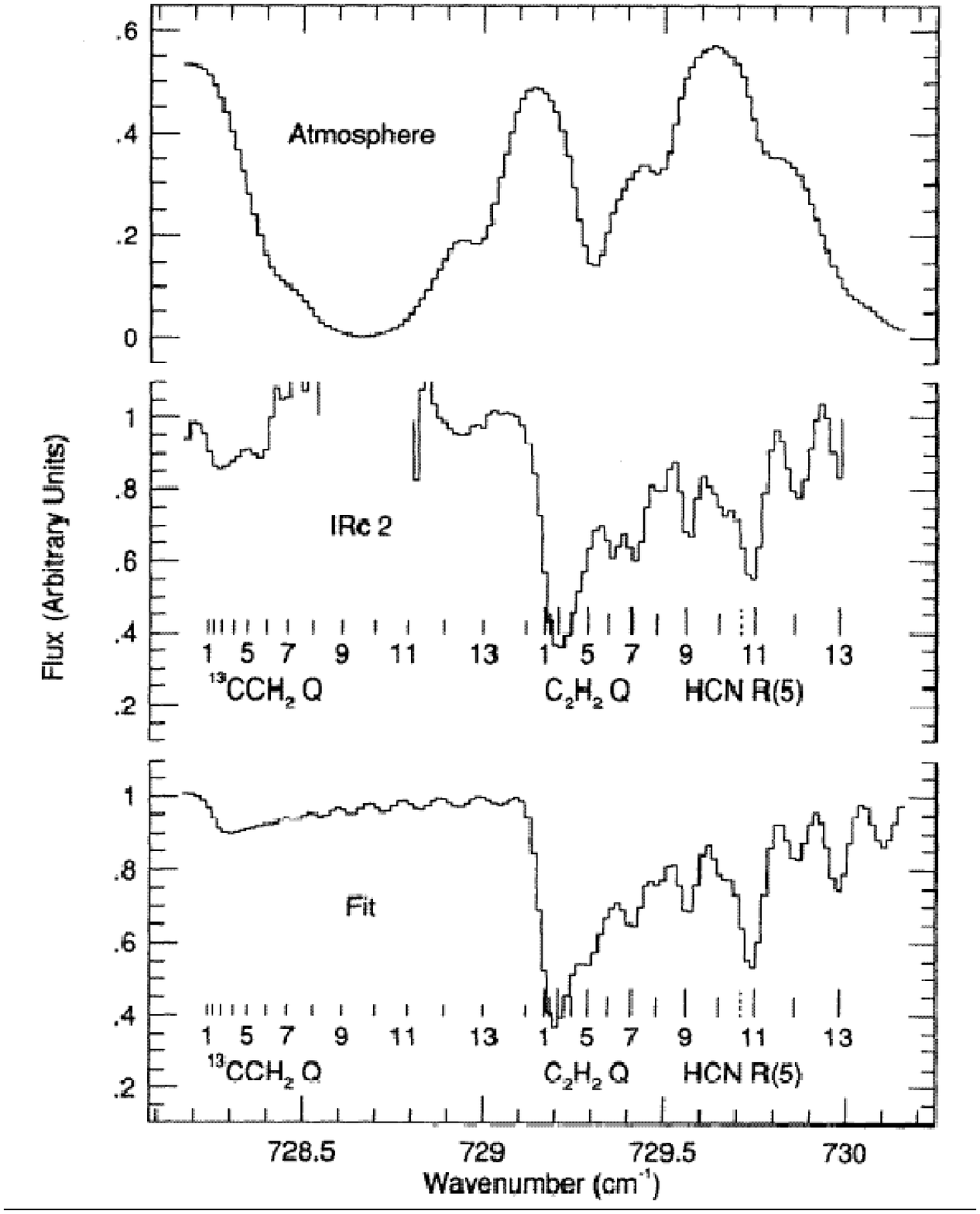}{3.in}{0}{30}{30}{-100}{0}
\caption{Molecular absorption toward IRc2 at the C$_2$H$_2$ Q-branch along
with observation of the atmosphere for comparison.  Note the number of lines
observed with a single setting.}
\end{figure}
These observations, taken from the ground with Irshell (Lacy et al. 1989),
demonstrate the information available with higher resolution and the
difficulty of observing through the Earth's atmosphere (note the top trace
showing an observation of the atmospheric absorption).

\section{High Spectral Resolution}

There are three major benefits to using high spectral resolution:
First, 
resolving the Earth's atmospheric lines makes observing between
the lines practical.  Furthermore, correcting for the atmospheric
lines becomes easier.  Second,
when the line width matches
the resolution of an instrument, the instrument's sensitivity to that
line is at a maximum.
Finally, if the line is fully resolved, line profile information becomes
available for modeling and analysis.  

Several types of instrument can achieve R=100,000 in the MIR, each with
advantages.  We will now discuss three methods to illustrate our 
decision in favor of 
a grating spectrograph.  We will not include coherent systems, such as
heterodyne spectrometers, which easily reach the resolution
requirement, but have limited sensitivity, spatial coverage, and 
spectral coverage (e.g. Betz 1980).

Fourier transform spectrometers (FTS), such as the one used at KPNO (Hall et al.
1979),
operate by sampling the Fourier transform of the spectrum over a range
of phase delays.  
For a more detailed
discussion of an FTS, see Hinkle (this volume).  
In the basic design, a Michelson interferometer, 
a beamsplitter divides the light between two separate
arms, the reference and the scanning arm,  and recombines before reaching
the detector.  
To achieve R=100,000
at 10~\micron, the scanning mirror must travel at least 0.5~m.  
The detector records the intensity as a function of the phase delay and
the resulting spectrum is recovered in software.  
If used with an array detector, an FTS will simultaneously sample the
spectrum of all points in the field-of-view (Maillard 1995).
Because the FTS samples
the entire bandpass at all times, the noise also comes from the entire 
bandpass and fluctuations in the sky background will affect the entire spectrum.
An FTS is most efficient when seeking extended wavelength coverage over
an extended object.

A Fabry-Perot (FP) is a scanning monochromator which measures spectral
elements sequentially.  With an array detector, it can also simultaneously
record spectra of an extended object, albeit with spectral shifts across
the field.  To achieve R=100,000 requires
multiple, cryogenic FP etalons.  Assuming a high plate finesse of
50, the spacing must be 1~cm and the free spectral range will be
0.5~cm$^{-1}$, requiring at least two additional elements, such as
a cryogenic FP and a narrow-band filter, to isolate the orders.
As with the FTS, an FP is subject to noise from sky fluctuations.
An FP is most efficient if observing a single line over an extended object.

A diffraction grating disperses light in a single dimension.  With a detector
array, a grating spectrograph will sample a continuous spectrum along with
one spatial dimension defined by the entrance slit.  To achieve R=100,000
at 10~\micron\ requires a grating $\geq$0.5~m long, as will be discussed below.
A grating is most efficient when observing a point source over an
extended spectral range.

\section{TEXES}

Because our science involves molecular
rotation-vibration lines, mostly in point sources, we chose to design
and build a grating spectrograph, the 
Texas Echelon-Cross-Echelle Spectrograph or TEXES.  TEXES operates
from 5~to~25~\micron\ with multiple spectroscopic modes: R$\approx$100,000 
in a cross-dispersed format; R$\approx$15,000 in single order;
and R$\approx$3,000 in single order.  
It is available to the community on a collaborative basis.
Recent observing runs at the McDonald 2.7~m telescope and
NASA's 3~m Infrared Telescope Facility (IRTF) have
demonstrated the capabilities of the instrument.  TEXES also serves as
a test-bed for construction of our SOFIA instrument, EXES.

The heart of TEXES is an echelon grating, a steeply blazed, coarsely ruled,
diffraction grating (Michelson 1898).
For a reflection grating, the grating
equation is
\begin{displaymath} 
m\,\lambda = 2\,d\,(sin\,\alpha + sin\,\beta)
\end{displaymath}
where $m$ is the grating order, $\lambda$ is the wavelength of interest, 
$d$ is the groove spacing, and $\alpha$ and $\beta$ are the angles of 
incidence and diffraction, respectively.
The theoretical resolving power in the diffraction limited case 
with $\alpha\approx\beta\equiv\theta$ is 
\begin{displaymath}
R={2\,L\,sin\,\theta\over\lambda}
\end{displaymath}
where $L$ is the length of the grating.
For high resolution within a confined volume, 
both the angle of incidence and the length of
the grating must be large.

Our echelon grating is a single piece of diamond-machined aluminum.
It is 36~inches long with a 3.4~inch square cross-section.  The angle
of incidence is 84.3\deg.  The groove spacing is 0.3~inch or, in more
familiar terms, 0.133 lines/mm.  
The grating was manufactured by Hyperfine, Inc
of Boulder, Colorado (Bach, Bach, \& Bach 2000).  
For images of the slightly longer EXES
grating, see 
\begin{displaymath}
http://nene.as.utexas.edu/exes/epo/figures/scifigs.html.
\end{displaymath}

The echelon design is very appropriate for a MIR instrument.  Since the
groove spacing is so coarse, we operate in 1500$^{th}$ 
order at 10~\micron\ with a limited free spectral range, 0.66214~cm$^{-1}$.
However, our MIR detector array, with only 256~$\times$~256~pixels, is 
small enough that orders overfill the array longward of 11~\micron.
In theory, the diffraction limited resolving power of our instrument is
R=180,000.  Because working at the diffraction limit is difficult to achieve
and inefficient due to light loss at the entrance slit, 
we typically operate at R$\approx$75,000.

We have conducted some laboratory tests to investigate the performance
of the echelon.  
Using a room temperature gas cell, we
create an emission line spectrum.  Figure 6 shows gas cell spectra
of CH$_4$ and C$_2$H$_2$ at 7.7~\micron\ and 13.7~\micron.  
\begin{figure}
\plotfiddle{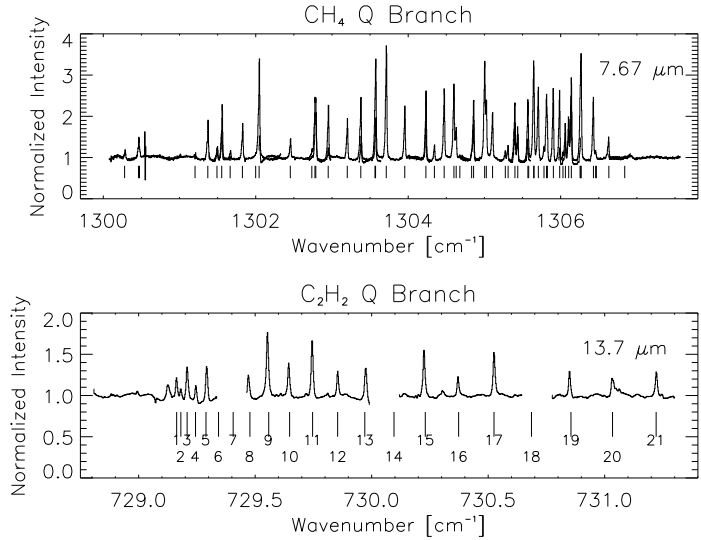}{2.75in}{0}{45}{45}{-150}{0}
\caption{Laboratory gas cell spectra of CH$_4$ and C$_2$H$_2$.
Each spectrum contains several orders that have been pieced together
in the 1D format.
Tick marks indicate the line frequencies.  
At the CH$_4$ setting, we have continuous wavelength coverage.
while at the
C$_2$H$_2$ setting, the orders are larger than the detector, causing
the gaps in the spectrum.}
\end{figure}
The C$_2$H$_2$ spectra
includes the region shown in Figure 5 observed with Irshell by Evans
et al. (1991).  With TEXES, it is easy to separate the individual
Q-branch lines.  From these spectra, we find deviations from
Gaussian line profiles
at 7.7~\micron\ and Gaussian profiles at 13.7~\micron\ (Figure 7).  
\begin{figure}
\plotfiddle{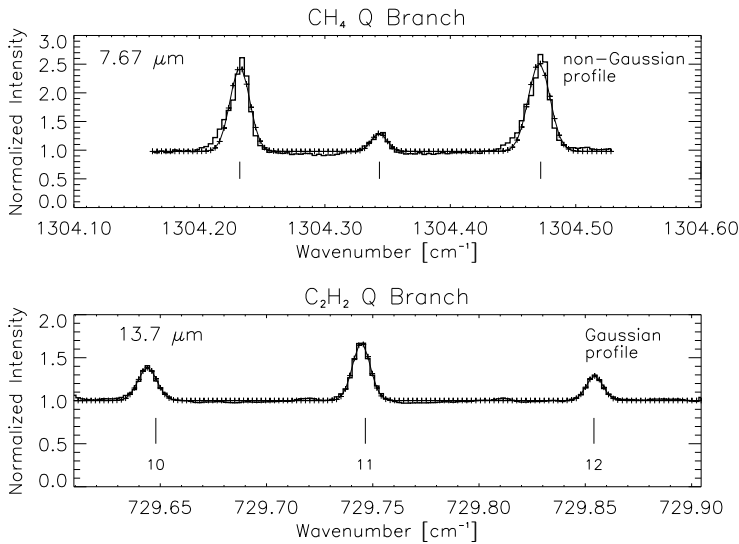}{2.75in}{0}{45}{45}{-150}{0}
\caption{The results of Gaussian fits to relatively isolated lines of
CH$_4$ and C$_2$H$_2$.  At the shorter wavelength, there is a wing to 
the spectral lines while the shape is very Gaussian at longer wavelengths.}
\end{figure}
Typical FWHM for each wavelength demonstrates R=75,000 in both cases.

TEXES's initial observing run was on the McDonald Observatory 2.7~m telescope
in February, 2000.  We had 8 nights for a combination of engineering and
science projects.  Figure 8 shows a high resolution, cross-dispersed
spectrum of Jupiter with numerous C$_2$H$_6$ lines in emission and
the same data summed along the slit and reorganized in a 1D format.
\begin{figure}
\plotfiddle{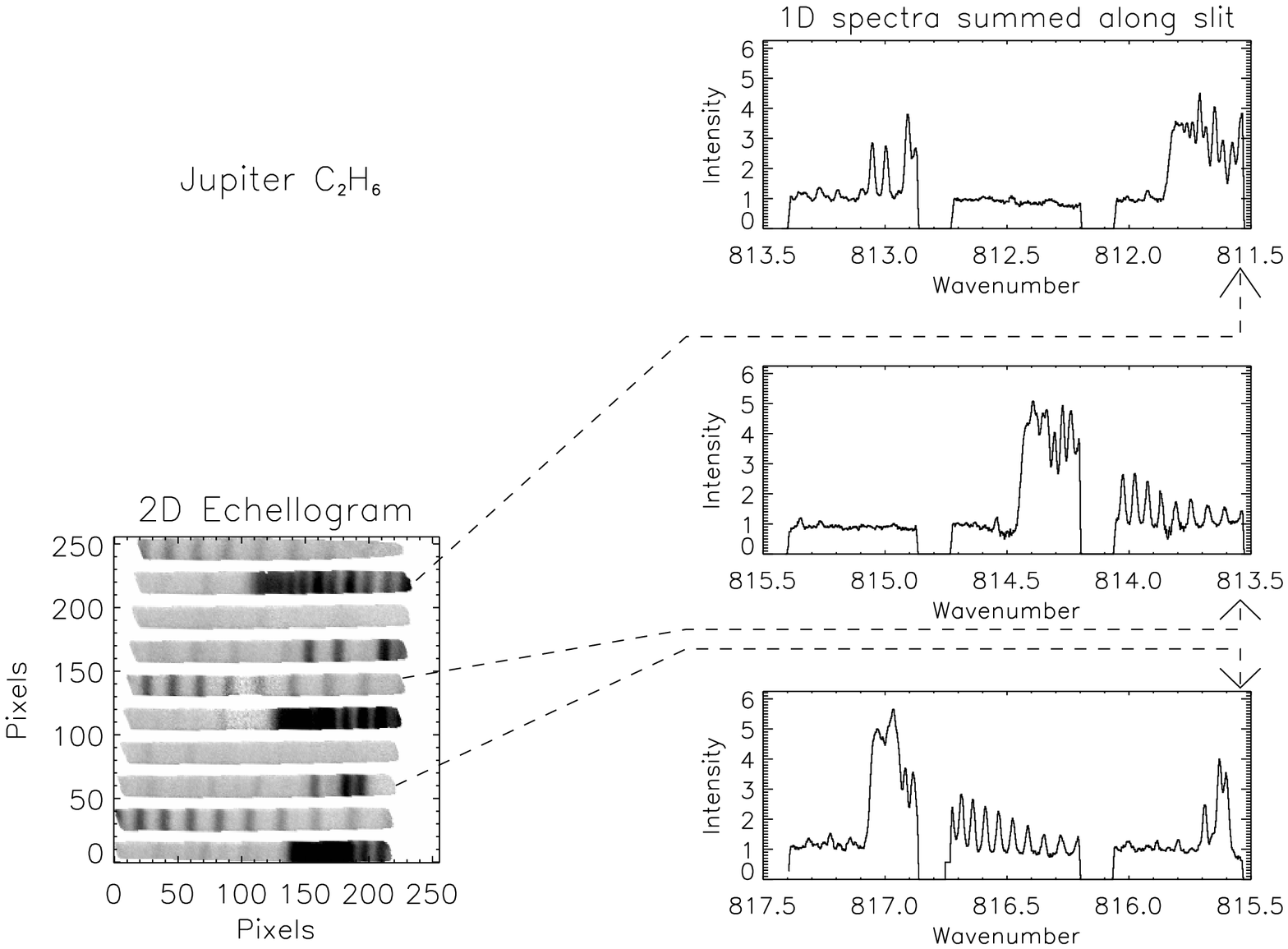}{4.75in}{0}{65}{65}{-225}{0}
\caption{A 2D echellogram showing C$_2$H$_6$ emission from Jupiter's 
stratosphere observed with TEXES at McDonald Observatory.  The integration
time was a few minutes on source with equal time on sky.  In the greyscale
image,
black represents the most intense emission, grey comes mostly from Jupiter's
continuum, and the unilluminated portions of the detector are white.  
The 1D plot shows the sum along the slit for each order with
three orders per panel.  In both cases, frequency increases down and
to the left (wavelength increases up and to the right).
The arrows indicate the relationship between
the two presentations.
At this wavelength, 12~\micron, orders are larger than the detector and
we do not get complete wavelength coverage, as seen by the regions where
intensity goes to zero.}
\end{figure}

We have subsequently had about 12 clear nights on NASA's IRTF.  With
a larger, infrared-optimized telescope at about twice the altitude, we
have had a significant improvement in our signal-to-noise of roughly a factor
of 10.  Projects
that were awarded time include: searching for H$_2$ pure rotational lines
in Uranus, planetary nebulae, and young stellar objects; examining C$_2$H$_2$
and HCN absorption toward massive star formation regions, looking at Jupiter
and Saturn in a variety of molecules, and investigating stellar Mg~I 
emission and H$_2$O absorption for a range of spectral types.  
Detailed results of these projects are awaiting complete data reduction
and will be published elsewhere, but
one preliminary observational result illustrates TEXES's performance.

Two Mg~I lines near 12~\micron\ (811.578 and 818.058~cm$^{-1}$) were found
in the Sun to be Zeeman sensitive (Brault \& Noyes 1983).  
These photospheric emission 
lines have been modeled by Carlsson, Rutten, \& Shchukina (1992) 
and are still used
for solar magnetic field studies (Moran et al. 2000).  Because the Zeeman
splitting increases with wavelength relative to Doppler broadening, these
MIR lines hold great potential for measurement of stellar magnetic fields.

Jennings et al. (1986) observed two stars, $\alpha$ Ori and $\alpha$ Tau,
with the KPNO 4~m telescope and FTS searching for the 811.578~cm$^{-1}$
line.  The resolving power selected was 50,000 with a wavelength coverage
of 2~cm$^{-1}$.  
Several hours of integration produced a firm detection in $\alpha$~Ori
and a probable detection in $\alpha$~Tau, both in absorption.  Later it
was recognized that the absorption in $\alpha$~Ori was from
H$_2$O (Jennings \& Sada 1998).

In Figure 9 we show a spectrum of $\alpha$~Tau obtained at the IRTF with
TEXES.  The spectral resolution is 75,000 (all lines resolved)
with a coverage of almost 6~cm$^{-1}$.  
This is a single nod pair with total
integration time of 4 seconds.
\begin{figure}
\plotfiddle{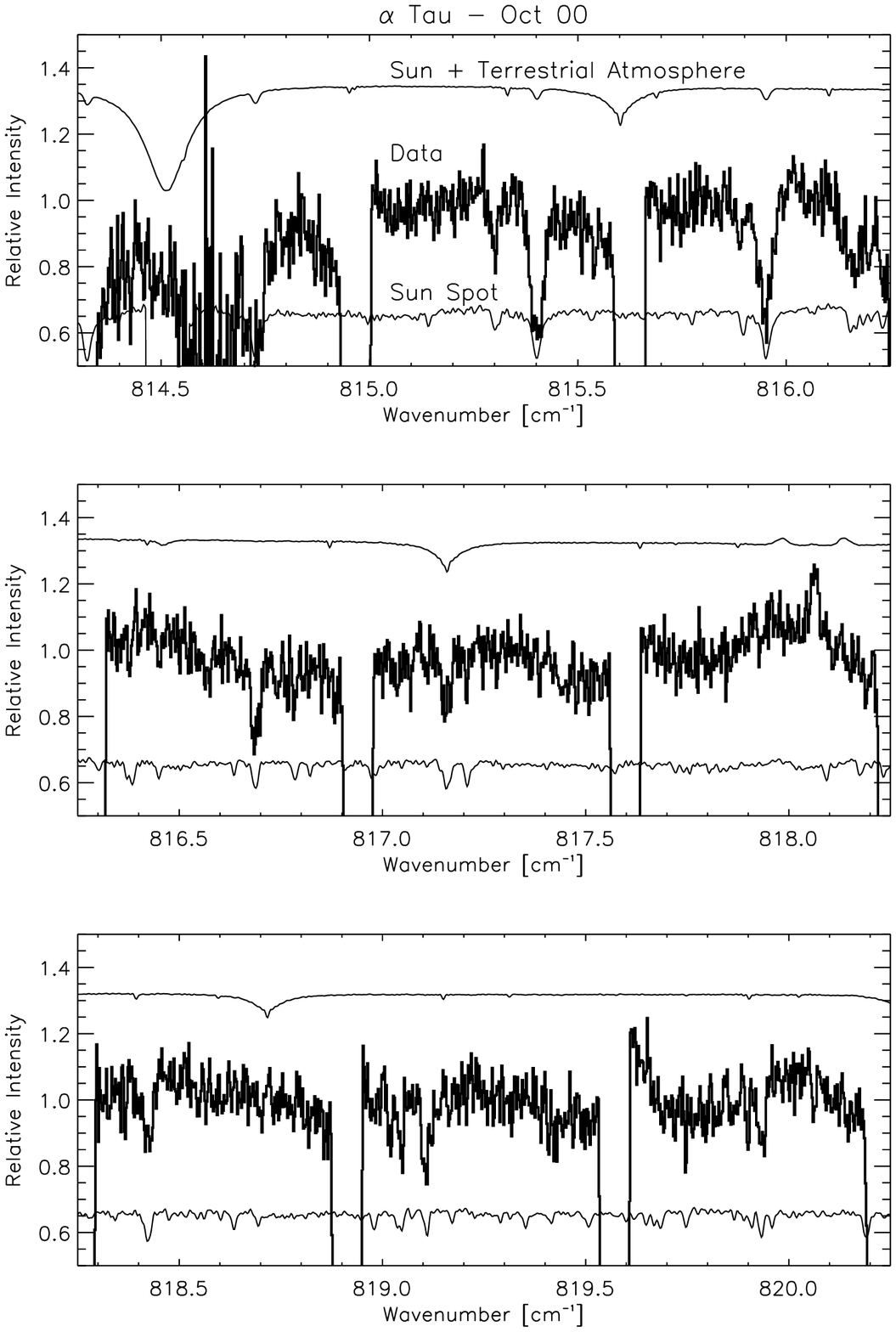}{5.75in}{0}{65}{65}{-155}{00}
\caption{TEXES spectrum of $\alpha$ Tau taken at the IRTF.  The
data (heavy line) are the result of differencing a single
image pair and using our pipeline data reduction program. 
For reference, we show a sunspot spectrum (lower line; offset by -0.35
in the figure) 
and a solar penumbra plus terrestrial atmosphere spectrum (upper line;
offset by +0.75)
taken from the KPNO Sunspot Atlas (Wallace,
Livingston, \& Bernath 1994).  In the penumbra spectrum, 
the Mg~I line at 818.06~cm$^{-1}$ is
completely split by the solar magnetic field and appears as two strong
components equally separated from a weaker one.
The $\alpha$ Tau data have been corrected for
velocity offsets.  The size of the shift
is seen near 814.5~cm$^{-1}$ where the increased noise in the data comes
from the terrestrial H$_2$O line seen in the upper trace. 
}
\end{figure}
The Mg~I line (818.06~cm$^{-1}$) is clearly seen in 
emission with numerous absorption features
coincident with OH and H$_2$O features identified in the sunspot spectrum.
Unlike the 811~cm$^{-1}$ line, this line does not 
suffer from blending with H$_2$O features.
From these data, if we assume a flux density 
for $\alpha$ Tau of 440~Jy (FD=416.7~Jy at 12.5~\micron; Hammersley
et al. 1998) our NEFD per spectral resolution element 
(1$\sigma$:1 second) is $\approx$10~Jy.

We have more time on the IRTF in June and will continue to make TEXES
available for collaborative projects.  Undoubtedly, some
important projects will require a larger aperture.
The foreoptics in TEXES make it adaptable to other telescopes and it
is our hope to someday be on a 8-10~m class telescope where our sensitivity
will improve by an order of magnitude.

\acknowledgments
Design, construction and observing with TEXES has been made possible
with the support of NSF grants AST-9120546 and AST-9618723, USRA grant
USRA 8500-98-008 and the Texas Advanced Research Project grant 
003658-0473-1999.

\end{document}